\documentclass[a4paper,12pt,english]{article}

\usepackage{amsfonts,bm,amssymb,euscript,array,babel,cite}

\usepackage[all]{xy}
\usepackage{amsmath,amsthm}
\usepackage[dvips]{graphicx}
\usepackage{color}
\usepackage[T1]{fontenc}

\makeatletter

\newcommand{\dd}{\mathrm{d}}

\newcommand{\w}{\wedge}
\newcommand{\bbm}{\left(\begin{matrix}}
\newcommand{\ebm}{\end{matrix}\right)}
\newcommand{\beq}{\begin{eqnarray}}
\newcommand{\eeq}{\end{eqnarray}}

\def\u{\upsilon}
\makeatother

\newcommand{\sfrac}[2]{{\textstyle\frac{#1}{#2}}}

\newcommand{\be}{\begin{equation}}
\newcommand{\ee}{\end{equation}}

\newcommand{\beqa}{\begin{eqnarray}}
\newcommand{\eeqa}{\end{eqnarray}} 
\def\nn{\nonumber} \def \bea{\begin{eqnarray}} \def\eea{\end{eqnarray}}

\newcommand{\barr}{\begin{array}}
\newcommand{\earr}{\end{array}}
\numberwithin{equation}{section}

 \def\d{\delta} 
    \def\k{\kappa}
\def\l{\lambda}   
 \def\o{\omega}   
 \def\S{\Sigma}

 \def\one{\mbox{1 \kern-.59em {\rm l}}}

\def\bit{\begin{itemize}} \def\eit{\end{itemize}}

\def\({\left(} \def\){\right)}

 \allowdisplaybreaks[3]

\begin{document}

\makeatother


\parindent=0cm

\renewcommand{\title}[1]{\vspace{10mm}\noindent{\Large{\bf

#1}}\vspace{8mm}} \newcommand{\authors}[1]{\noindent{\large

#1}\vspace{5mm}} \newcommand{\address}[1]{{\itshape #1\vspace{2mm}}}


\begin{titlepage}

\begin{flushright}
{ITP-UH-16/15}
\end{flushright}

\begin{center}

\vskip 3mm

\title{ {\Large
T-duality without isometry via 
extended 
gauge symmetries of 2D sigma models
} }

\vskip 3mm

 \authors{\large 
 Athanasios {Chatzistavrakidis}$^{\sharp,\circ,}\footnote{a.chatzistavrakidis@rug.nl}$, Andreas Deser$^{\sharp,}\footnote{andreas.deser@itp.uni-hannover.de}$, Larisa Jonke$^{\sharp,\dagger,}\footnote{larisa@irb.hr}$ }

\vskip 3mm

\address{ {$^{\sharp}$}\
Institut f\"ur Theoretische Physik,  
Leibniz Universit\"at Hannover,\\  Appelstra{\ss}e 2, 30167 Hannover, Germany 

\

  {$^{\circ}$}\ Van Swinderen Institute for Particle Physics and Gravity, University of Groningen, \\ 
        Nijenborgh 4, 9747 AG Groningen, 
        The Netherlands

\

{ {$^{\dagger}$}\
 Division of  Theoretical Physics, 
 Rudjer Bo$\check s$kovi\'c Institute, \\
 Bijeni$\check c$ka 54, 10000  Zagreb, Croatia}
}


\end{center}

\vskip 2cm

 \begin{center}
\textbf{Abstract}

 \end{center}
\vskip 3mm


Target space duality is one of the most profound properties of string theory. However it customarily requires that the background fields 
 satisfy certain invariance conditions in order to perform it consistently;
  for instance
  the vector fields along 
the directions that T-duality is performed have to generate isometries. In the present paper we examine in detail the 
possibility to perform T-duality along non-isometric directions. 
In particular, 
based on a recent work of Kotov and Strobl, 
 we study gauged 2D sigma models where gauge 
invariance for an extended set of gauge transformations
 imposes weaker constraints than in the standard case, notably the corresponding 
 vector fields are not Killing. 
 This formulation enables us to follow a procedure analogous to the derivation 
of the Buscher rules and obtain two dual models, by integrating out 
once the Lagrange multipliers and once the gauge fields. We show that this construction 
indeed works in non-trivial cases by examining an explicit class of examples based on step 2 nilmanifolds.

%

\end{titlepage}

\tableofcontents

\section{Introduction}

Dualities play a prominent role in many corners of modern theoretical physics (see Ref. \cite{Polchinski:2014mva} for a 
very interesting recent discussion). In string theory dualities are instrumental in 
understanding the structure of the theory and study its fundamental properties. 
Notably, T-duality \cite{Giveon:1994fu} is a symmetry of string theory that relates compactified 
backgrounds with inverse radii. The string background fields transform under T-duality 
according to a set of rules determined by Buscher in the seminal papers \cite{Buscher1,Buscher2}. This 
symmetry was subsequently proven a true symmetry between conformal field theories in Ref. \cite{RocekVerlinde}.

The approach followed by Buscher requires the existence of global isometries in the 2D sigma model, which are subsequently gauged. This approach was also followed in more involved cases, such as when there 
is a non-Abelian set of vector fields \cite{HullSpence1,HullSpence2,Hull,Alvarez1,Alvarez2,Plauschinn1,Plauschinn2,delaOssa}. 
In all cases there is a set of invariance conditions and constraints to be obeyed. 
Most importantly the 
vector fields generate isometries, which seems to be necessary in order to write down a sigma model 
that is gauge invariant under standard gauge transformations.

Recently a new twist appeared in the construction of gauged   sigma models. Kotov and Strobl (KS) \cite{Kotov:2014iha} proved the existence of gauged symmetries in sigma models which do not correspond to global 
 ones. This formulation is based on gauge symmetries associated to Lie algebroids{\footnote{For the purposes of this paper it will be sufficient to think of Lie algebroids simply as a generalization of Lie algebras with 
 $X$-dependent structure functions instead of structure constants for a bracket that satisfies the 
 Jacobi identity.}}, 
 a field pioneered by Strobl in the context of Yang-Mills theories \cite{Strobl:2004im} and 
 studied further in Refs. \cite{Bojowald:2004wu,Strobl1,Strobl2,Strobl3,Mayer}. 
 The essence of this formulation relies on an extension of the standard infinitesimal gauge transformations 
 for the gauge fields $A^a$ to include a part proportional to $DX^i$, where $X^i$ are the world sheet scalars 
 and $DX^i$ is the gauge covariant derivative on the world sheet obtained by minimal coupling. Then it is possible to 
 construct an action which is invariant under these extended gauge transformations upon a milder 
 condition than isometry. We will explain the basics of this formulation in section 2.
 
The above remarkable result immediately indicates that one can revisit Buscher's procedure in a 
more general context where isometries are not present. Earlier attempts to understand dualities on general backgrounds include Poisson-Lie T-duality \cite{Klimcik:1995dy,Klimcik:1995jn,Klimcik:1995ux,Sfetsos:1997pi},  generalized T-duality for cases with no globally defined Killing vectors \cite{Hull}, and an approach
based on ``covariant coordinates'' \cite{Davidovic:2015mpa}.  In the string theory 
context this is an important problem, given that one often encounters backgrounds that do not have isometries but one 
would like to know their T-dual backgrounds. In this paper we take this challenge. In particular 
we employ the formulation of KS and study gauged sigma models without isometry. These include 
a set of gauge fields as well as Lagrange multipliers in the same spirit as in Buscher's procedure. 
The conditions and constraints to be obeyed are determined and shown to be milder than the isometric case. 
This allows us to obtain two dual models, one by integrating out the Lagrange multipliers thus obtaining 
the original ungauged model, and one by integrating out the gauge fields. The latter yields a 
dual model which we describe in precise terms.

Given that several constraints appear in the formulation, it is natural to worry whether any non-trivial 
cases exist at all, namely whether the formulation is empty of non-trivial examples and isometry 
is always restored. We prove by an explicit toy example that this is not the case. This example 
is based on a well-known manifold used in studies of string duality, the 3D Heisenberg nilmanifold. 
This is a parallelizable manifold with a global section of its tangent bundle. The vector fields that 
form a basis for any such section are known and while one of them is Killing, the other two are 
not. Nevertheless, in the formulation established in the present paper we are able to T-dualize along 
all three directions. We perform this procedure in detail and discuss the dual model. 
Furthermore, we show that this is not an isolated example; the full class of step 2 nilmanifolds can be treated the same way, as we show in section 5. The examples we examine in this paper are not proper string backgrounds as they are not conformal; however they are often discussed in literature since they appear as T-dual of tori with $H$ flux and have simple yet non-trivial geometric description.

 \section{Action and gauge symmetry} 

 \subsection{Preliminaries}
 
 Let us consider the standard $\sigma$-model action for the bosonic sector of closed string theory at leading order in $\alpha'$,
  \bea\label{isomodel}
 S&=&\int_{\Sigma_2}\sfrac 12 g_{ij}\dd X^i\wedge \star \dd X^j+\int_{\Sigma_3} \sfrac 16 H_{ijk}\dd X^i\wedge\dd X^j\wedge\dd X^k~,
 \eea
 where $\Sigma_2=\partial\Sigma_3$ is the 2D world sheet and $X=(X^i):\Sigma_2\to \text{M}$ is the map from the 
 world sheet to the target space $\text{M}$. Here and in the following we ignore the dilaton coupling, which enters 
 the action at linear order in $\alpha'$ and leave a discussion on this issue for the future, since it requires different techniques than the ones we introduce here. 
 
 The standard approach to T-duality begins with the assumption of \emph{global} target space symmetries generated 
 by vector fields $\upsilon_a=\upsilon_a^i\partial_i$. This means that the action is required to be 
 invariant under the 
 global transformations
 \be 
 \d_{\epsilon}X^i=\upsilon^i_a(X)\epsilon^a~,
 \ee
 where $\epsilon^a$ are rigid transformation parameters. It is well-known that the  invariance of the action \eqref{isomodel} is not automatic 
 but imposes the constraints 
 \bea
 {\cal L}_{\upsilon_a}g&=&0~,\label{iso1}\\
 \iota_{\upsilon_a}H&=&\dd \theta_a~,\label{iso2}
 \eea
 for some arbitrary 1-forms $\theta_a=\theta_{ai}\dd X^i$. This is true regardless whether the vector fields 
 generate an Abelian or a non-Abelian algebra. We will let them here satisfy a non-Abelian one with 
 structure constants $C^c_{ab}$, 
 \be 
 [\upsilon_a,\upsilon_b]=C_{ab}^c\upsilon_c~.
 \ee
 The next step is to gauge the above global symmetry. This is performed via the usual minimal coupling to gauge fields 
 (1-forms) $A^a$, where the de Rham differentials on the world sheet are substituted by 
 \be 
 DX^i=\dd X^i-\upsilon^i_a(X)A^a~,
 \ee
 and the local (gauge) transformations are given as 
 \bea 
 \d_{\epsilon}X^i&=&\upsilon^i_a(X)\epsilon^a(X)~,\nn\\
 \d_{\epsilon}A^a&=&\dd \epsilon^a(X)+C^a_{bc}A^b\epsilon^c(X)~,
 \eea
 with $\epsilon^a=\epsilon^a(X)$ the gauge parameters. The corresponding gauged action 
\cite{HullSpence1} includes additional fields but we will not 
 discuss its precise form yet because we are going to present a more general result below. However let us mention that 
 gauge invariance imposes additional constraints on top of \eqref{iso1} and \eqref{iso2}. All these conditions and 
 constraints will appear as a certain limit of the more general formulation that we present immediately below. 
 
 \subsection{Gauging without isometry}
 
 As described in Ref. \cite{Kotov:2014iha}, it is possible to write down gauged 2D $\sigma$-models even when there is no 
 isometry to begin with. This is rather unconventional from a standard gauge theory viewpoint, 
 where normally we gauge a symmetry 
 that is already there as a rigid one. Here we refer to local symmetries that do not possess a global counterpart. 
 
To be precise, let us consider the gauged action
 \bea\label{Sg}
 S&=&\int_{\Sigma_2}\sfrac 12 g_{ij}DX^i\wedge \star DX^j
 +\int_{\Sigma_3} \sfrac 16 H_{ijk}\dd X^i\wedge\dd X^j\wedge\dd X^k-\nn\\
 &-& \int_{\Sigma_2}(\theta_a+\dd\eta_a)\wedge A^a+\int_{\Sigma_2}\sfrac 12(\iota_{\u_{[a}}\theta_{b]}+C^c_{ab}\eta_c)A^a\wedge A^b
 -\int_{\Sigma_2}\omega^a_{bi}\eta_a A^b\wedge DX^i~.
 \eea
 The explanation of the ingredients is as follows. 
First of all, we defined the 1-forms
\bea
&& DX^i=\dd X^i-\u^i_a A^a,\nn\\
&&   \theta_a=\theta_{ai}\dd X^i,
\eea
where $A^a$ are again the gauge fields, and a set of auxiliary scalar fields $\eta_a$. The vector fields $\u_a$ 
satisfy $$[\u_a,\u_b]=C^c_{ab}(X)\u_c~,$$
where now the structure functions $C_{ab}^c$ are not necessarily constants, namely we allow them to depend on $X^i$. This provides a straightforward generalization to sections in arbitrary Lie algebroids{\footnote{We point out that this is a \emph{possible} generalization which is implemented here, but not a \emph{necessary} one. 
The formulation we present is already a generalization of the standard one even in the Lie algebra case.}}. Finally, $\omega^a_{bi}$ are the components of 
a connection 1-form $\omega^a_b=\omega^a_{bi}\dd X^i$ that twines the spacetime indices with the gauge ones. Its role will be clarified immediately below. Note that for vanishing $\omega^a_{bi}$ the action \eqref{Sg} is precisely the one considered in Refs. \cite{HullSpence1,Plauschinn2}, where 
T-duality \emph{with} isometry was studied{\footnote{In Ref. \cite{HullSpence1} the fields $\eta_a$ do not play a crucial role. 
This was revisited in Ref. \cite{Plauschinn2} where these fields are present and transform 
non-trivially under gauge transformations. This will be the case in our formulation too.}}. 
The geometric interpretation of $\omega$ as a connection 1-form was first introduced in \cite{Mayer} (see also \cite{Strobltoappear}); one can then introduce the corresponding exterior covariant derivative 
\beq D^\omega = d + \omega \wedge\eeq
 and curvature
\beq R^a_b = D^\omega \omega^a_b = d\omega^a_b + \omega^a_c\wedge \omega^c_b\;.\eeq 
The transformation properties of $\omega^a_{bi}$ are the same as for the spin connection. 
In particular, being an 1-form, it transforms covariantly in the index $i$.

Let us now specify the gauge transformations for the fields $X^i,A^a$ and $\eta_a$. These have the form
\bea\label{gt}
&& \delta_\epsilon X^i=\u^i_a\epsilon^a~,\nn\\
&& \delta_\epsilon A^a=\dd \epsilon^a+C^a_{bc}A^b\epsilon^c+\omega^a_{bi}\epsilon^bDX^i~,\nn\\
&& \delta_\epsilon \eta_a=-\iota_{\u_{(a}}\theta_{b)}\epsilon^b-C^c_{ab}\epsilon^b\eta_c+\u^i_a\omega^d_{bi}\eta_d\epsilon^b~,
\eea
for $X$-dependent parameters $\epsilon^a(X)$.
It is directly observed that the gauge transformation for the gauge field $A^a$ is extended 
in comparison to the standard one by 
an $\omega$-dependent term proportional to $DX^i$. This is a key ingredient of the present formulation. Note that in 2D one can add a term proportional to $\star DX^i$; this is currently under investigation~\cite{CDJS}.

The action (\ref{Sg}) is invariant under the above $\omega$-extended gauge transformations
provided that the following conditions hold
\bea\label{noiso1}
&& {\cal L}_{{\u}_a}g=\o^b_a\vee\iota_{{\u}_b}g~,\\
\label{noiso2}
&& \iota_{\u_a} H=\dd\theta_a+\theta_b\w\omega^b_a -\eta_b R^b_a~,
\eea
where $\vee$ denotes the symmetric product{\footnote{This means that Eq. \eqref{noiso1} reads in components 
as $({\cal L}_{{\u}_a}g)_{ij}=2\o^b_{a(i}\u_{\underline{b}}^kg_{j)k}$, where the symmetrization is weighted, and it is obviously covariant, since 
$\omega^b_{ai}$ is an 1-form.}}.
Similarly to the standard case there is a set of additional constraints, 
which now become 
\bea\label{constraint1}
&&{\cal L}_{{\u}_{[a}}\theta_{b]}= C^d_{ab}\theta_d-\iota_{\u_d}\theta_{[a}\omega^d_{b]}-\iota_{\u_{[a}}\omega^d_{b]}\theta_d -D^c_{ab} \eta_c~,
\\
\label{constraint2}
&& \sfrac 13 \iota_{\u_a}\iota_{\u_b}\iota_{\u_c}H=\iota_{\u_{[a}}C^d_{bc]}\theta_d-2\iota_{\u_{[a}}\omega^d_b \iota_{\u_{c]}}\theta_d
-2{\tilde D}^e_{abc} \eta_e~,\eea
where we defined the shorthand notation
\bea
D^e_{ab}&=&\dd C^e_{ab}+C^c_{ab}\omega^e_c+2C^e_{d[a}\omega^d_{b]}
+2\iota_{\u_d}\omega^e_{[b}\omega^d_{a]}+2{\cal L}_{\u_{[b}}\omega^e_{a]}
+\iota_{\u_{[a}}R^e_{b]}~,\nn\\
{\tilde D}^e_{abc}&=&\iota_{\u_{[a}}\iota_{\u_b}R^e_{c]}~.
\eea
This result is obtained using the identity 
\be 
C^d_{[ab}C^e_{c]d}+\u^k_{[c}\partial_{\underline{k}}C^e_{ab]}=0~,
\ee
which is the Jacobi identity in the Lie algebroid case where the structure functions are not constant.
 Note that sending $\omega^a_{bi}$ to zero and the functions $C^a_{bc}$ to constants restores the isometric case and 
 all the conditions fully agree{\footnote{Note that our conventions 
 and notation are slightly different.}} with the results of Ref. \cite{Plauschinn2}. 
 One apparent difference between our formulation and previously studied ones is the explicit dependence of the 
 constraints \eqref{constraint1} and \eqref{constraint2} on the scalar fields $\eta_a$. These scalar fields are essentially 
 the analogues of the Lagrange multipliers introduced in Buscher's procedure, which become the coordinates of the T-dual model 
 upon integration of the gauge fields.

At this stage it is useful to discuss the field strength of the gauge fields $A^a$. Recall that 
the 2-form that multiplies the Lagrange multipliers in Buscher's procedure is precisely the 
field strength of the corresponding gauge fields. In the present formulation this turns out to be 
\be 
{\cal F}^a := \dd A^a+\sfrac 12 C^a_{bc}A^b\w A^c-\omega^a_{bi}A^b\w DX^i~,
\ee  
which is the same as the one considered in Ref. \cite{Mayer}. 
A straightforward calculation confirms the result of \cite{Mayer} on the gauge transformation 
of this field strength:
\be 
\d_{\epsilon}{\cal F}^a=(C^a_{bc}-\omega^a_{ci}\u^i_b)\epsilon^c{\cal F}^b+R^a_{bij}\epsilon^bDX^i\w DX^j+
D^{a}_{bci}\epsilon^c DX^i\w A^b~,
\ee 
Although in the present paper we do not consider dynamics for the gauge fields, the above transformation 
rule is very suggestive. A covariant transformation rule for the field strength ${\cal F}^a$ requires 
\beq
R^a_b=0 \quad \text{and} \quad D^a_{bc}=0~,
\eeq
namely the flatness of the connection $\omega$. In that case it is immediately observed 
that the $\eta$ dependence in the constraints \eqref{constraint1} and \eqref{constraint2} drops out. 
This will be the case in the explicit examples that will be presented in later sections, where we will also
make some essential comments about this flatness condition.

\section{T-duality}

In the previous section we considered the gauged action for a $\sigma$-model and discussed under which conditions it is 
gauge invariant. Now we would like to follow the spirit of Buscher's approach to T-duality and obtain the two 
T-dual models that stem from this action. In order to do so, we have to integrate out two different sets of fields. 
The original model should be obtained upon integration of the Lagrange multipliers $\eta_a$ 
and gauge fixing, while the dual model 
is obtained by integrating out the gauge fields $A^a$. 

\subsection{Recovering the ungauged model}

In order to recover the ungauged original model \eqref{isomodel} we follow the steps described in detail below.
First
we lift the full action modulo the kinetic term to three dimensions{\footnote{Using differentiation and Stoke's theorem, and ignoring possible global issues.}:
\bea\label{Sg33}
 S&=&\int_{\S_2}\sfrac 12 g_{ij}DX^i\wedge \star DX^j+\int_{\S_3} \sfrac 16 H_{ijk}\dd X^i\wedge\dd X^j\wedge\dd X^k-\nn\\
 &-& \int_{\S_3}\dd(\theta_a+\dd\eta_a- \omega^b_a\eta_b)\wedge A^a+\int_{\S_3}(\theta_a+\dd\eta_a- \omega^b_a\eta_b)\wedge\dd A^a+\nn\\
 &+&\int_{\S_3}\sfrac 12\dd(\iota_{\u_{[a}}\theta_{b]}+C^c_{ab}\eta_c-2\omega^c_{[b\underline{i}}\u^i_{a]}\eta_c)\w A^a\wedge A^b-\nn\\
 &-&\int_{\S_3}(\iota_{\u_{[a}}
 \theta_{b]}+C^c_{ab}\eta_c-2\omega^c_{[b\underline{i}}\u^i_{a]}\eta_c)A^a\wedge\dd A^b
 .
 \eea
Next, we covariantize the de Rham differentials of the $H$ term and obtain:
\bea\label{Sg33c}
 S&=&\int_{\S_2}\sfrac 12 g_{ij}DX^i\wedge \star DX^j+\int_{\S_3} \sfrac 16 H_{ijk}DX^i\wedge DX^j\wedge DX^k+\nn\\
 &+& \int_{\S_3}\left( \sfrac 12 H_{ijk}\u^i_a A^a\w\dd X^j\w\dd X^k - \sfrac 12 H_{ijk}\u^i_a \u^j_bA^a\w A^b\w\dd X^k+
  \sfrac 16 H_{ijk}\u^i_a \u^j_b\u^k_cA^a\w A^b\w A^c\right)-\nn\\
 &-& \int_{\S_3}\dd(\theta_a+\dd\eta_a- \omega^b_a\eta_b)\wedge A^a+\int_{\S_3}(\theta_a+\dd\eta_a- \omega^b_a\eta_b)\wedge\dd A^a+\nn\\
 &+&\int_{\S_3}\sfrac 12\dd(\iota_{\u_{[a}}\theta_{b]}+C^c_{ab}\eta_c-2\omega^c_{[b\underline{i}}\u^i_{a]}\eta_c)\w A^a\wedge A^b-
 \int_{\S_3}(\iota_{\u_{[a}}\theta_{b]}+C^c_{ab}\eta_c-2\omega^c_{[b\underline{i}}\u^i_{a]}\eta_c)A^a\wedge\dd A^b
 \nn.
 \eea
Using the constraints imposed by gauge invariance and after a long and tedious calculation the action can be written in the following form:
\bea\label{Sgeom}
 S&=&\int_{\S_2}\sfrac 12 g_{ij}DX^i\wedge \star DX^j+\int_{\S_3} \sfrac 16 H_{ijk}DX^i\wedge DX^j\wedge DX^k+\nn\\
 &+& \int_{\S_3}(\theta_a-\iota_{\u_{[b}}\theta_{a]}A^b)\w(\dd A^a+\sfrac 12 C^a_{bc}A^b\w A^c-\omega^a_{bi}A^b\w DX^i)+\\
 &+&\int_{\S_3}(\dd\eta_a-C^c_{ba}\eta_c A^b-\omega^b_a\eta_b+2\omega^c_{[a\underline{i}}\u^i_{b]}\eta_c A^b)\w(\dd A^a+\sfrac 12 C^a_{bc}A^b\w A^c
 -\omega^a_{bi}A^b\w DX^i)~.\nn
 \eea
Now we integrate the Lagrange multiplier $\eta_a$ from the gauged action. The equation of motion for $\eta_a$ is:
\bea\label{eta}
{\cal F}^a = \dd A^a+\sfrac 12 C^a_{bc}A^b\w A^c-\omega^a_{bi}A^b\w DX^i=0.
\eea
This is the field strength we discussed in the previous section, which is simply the standard $F^a$ of the non-Abelian gauge 
fields $A^a$ when $\omega^a_{bi}=0$. 
Inserting (\ref{eta}) in the form  (\ref{Sgeom}) of the action gives
 \bea\label{Sung}
S&=&\int_{\S_2}\sfrac 12 g_{ij}DX^i\wedge \star DX^j+\int_{\S_3} \sfrac 16 H_{ijk}DX^i\wedge DX^j\wedge DX^k~.
\eea
This result is identified with the original model in the same spirit as in the (Abelian or non-Abelian) 
isometric case \cite{RocekVerlinde,delaOssa}. In particular, since ${\cal F}^a=0$ the gauge fields must be 
pure gauges. Since at this stage we are working on-shell, these gauges may be fixed. The simplest 
gauge choice, which is the same as the one that was considered in Refs. \cite{RocekVerlinde,delaOssa}, is $A^a=0$. 
Then one immediately recovers the original model. Different gauge choices are of course allowed too, 
and then the original model in different coordinate systems is recovered{\footnote{Note that since the metric depends on $X^i$ this procedure has to be carefully performed. We will provide a detailed account on that 
in a class of examples later on.}}. 

Finally let us note that introducing the shorthand notation\footnote{Neither the structure constants $C^a_{bc}$ nor the connection $\omega^a_b$ transform as tensors, but the combination $K^a_{bc}$ does.} 
\bea
K^c_{ab} = 2\iota_{\u_{[b}}\omega^c_{a]} + C^c_{ab}~,\eea
 the action \eqref{Sgeom} can be rewritten in the form
 \bea\label{SgeomHull}
 S&=&\int_{\S_2}\sfrac 12 g_{ij}DX^i\wedge \star DX^j+\int_{\S_3} \sfrac 16 H_{ijk}DX^i\wedge DX^j\wedge DX^k+\nn\\
 &+&\int_{\S_3}\,{\cal F}^a \wedge \theta_{ai} DX^i 
 + \int_{\S_3} \, {\cal F}^a \wedge (D^\omega \eta_a + K^c{}_{ab}\eta_c A^b)~.
 \eea
Comparing with the Abelian, isometric case (i.e. setting $\omega$ and $C^a_{bc}$ to zero) and disregarding $\eta_a$ we recover the action given earlier in the literature, e.g. \cite{Hull}. Thus \eqref{SgeomHull} is a natural generalization thereof, obtained by replacing $\dd A^a$ by the appropriate field strength ${\cal F}^a$ and introducing the auxiliaries $\eta_a$ in a covariant way. Moreover, since Eq. \eqref{Sgeom} gives our starting action \eqref{Sg}, this serves as an additional geometric motivation for the 
introduction of the $\omega$- and $\eta$-dependent terms.

 \subsection{Obtaining the dual model}
 
 Let us now turn our attention to the dual model. 
 This is obtained by integrating out the gauge fields $A^a$ from the action.
 Varying the action (\ref{Sg}) with respect to $A^a$ results in the equations of motion 
 \bea\label{eomA}
 -g_{ij}\u^i_a\star\dd X^j+g_{ij}\u^i_a\u^j_b\star A^b+(\iota_{\u_{[a}}\theta_{b]}+K^c_{ab}\eta_c)A^b+\theta_a+D^\omega\eta_a =0~.
 \eea
 Similarly to \cite{Hull,Plauschinn2}, it is useful to define the following tensors:
 \bea 
 G_{ab}&=&\u^i_a g_{ij}\u^j_b~,\label{Gab}\\
 D_{ab}&=&\iota_{\u_{[a}}\theta_{b]}+K^c_{ab}\eta_c~,\label{Dab}
 \eea
 and 
 \bea 
 \xi_a&=& \theta_a+D^\omega \eta_a~,\\
 \u^*_a&=&g_{ij}\u^i_a\dd X^j~,
 \eea
 which now contain the components of $\omega$.
 The equation of motion takes the simpler form
\bea\label{eomAa}\star \u^*_a-\xi_a=G_{ab}\star A^b+D_{ab}A^b~.\eea
Inserting this into the action (\ref{Sg}) yields an expression linear in the gauge fields: 
 \bea\label{adual} S=\int_{\S_2} \sfrac 12 g_{ij}\dd X^i\w\star\dd X^j-\int_{\S_2} \sfrac 12 A^a\w (\star \u^*_a-\xi_a) +\int_{\S_3}\sfrac 16 H_{ijk}\dd X^i\w\dd X^j\w\dd X^k~.\eea
 The next step requires solving  equation (\ref{eomAa}). 
 This can be done as follows. We make the general Ansatz
 \be 
 A^a=M^{ab}\u^{\star}_b+N^{ab}\xi_b+P^{ab}\star\u^{\star}_b+Q^{ab}\star\xi_b~,
 \ee 
 with coefficients to be determined. 
 Inserting this Ansatz in the equation of motion and using $\star^2=1$   we obtain the matrix equations
 \bea 
 GM+DP&=&1~,\nn\\
 GN+DQ&=&0~,\nn\\
 GP+DM&=&0~,\nn\\
 GQ+DN&=&-1~,
 \eea 
 with $G$ and $D$ given in \eqref{Gab} and \eqref{Dab}. The solution of this system 
 gives 
 \be 
 Q=-(G-DG^{-1}D)^{-1}~,
 \ee
 and the rest of the unknowns are determined in terms of $Q$ as 
 \bea 
M&=&-Q~,\nn\\
N&=&-G^{-1}DQ~,\nn\\
P&=&G^{-1}DQ~.
 \eea
 Then $A^a$ is determined and may be inserted in the action.
The resulting action of the dual model is 
\be 
S=\int_{\S_2}\left(\sfrac 12 (G-DG^{-1}D)^{ab}e_a\w\star e_b
-\sfrac 12\big(G^{-1}D(G-DG^{-1}D)^{-1}\big)^{ab}e_a\w e_b \right)+\int_{\S_3}H~,
\ee
where 
\be 
e_a=\dd\eta_a+\theta_a-(\omega^b_{ai}\eta_b+(G^{-1}D)^b_a\u^k_bg_{ki})\dd X^i~.
\ee 
We observe that the dual action comprises a coframe that mixes the original coframe with the 
dual one. Thus the generic result is that the coordinates of the original and the dual model appear 
mixed and cannot always be disentangled. We will have more to say about this in the following section, where 
we study a non-trivial example.

\section{Example - 3d nilmanifold}

In the previous section we presented the formulation that leads to two dual models in the absence of isometry. 
Evidently this depends crucially on the connection 1-form coefficients $\omega^a_{bi}$. In particular, when these 
coefficients vanish isometry is restored. Therefore in order to be able to argue that this formulation 
is not an empty and useless theoretical method it is necessary to show that it works in non-trivial cases. 
This is not obvious, given that a lot of constraints were imposed and thus one might worry that they do not allow 
for non-vanishing $\omega^a_{bi}$. 
 In the present section we work out 
an explicit example which serves as an existence proof and supports the non-triviality of our considerations. 
In the next section we discuss a larger class of examples.

\subsection{The background and the action}

Let us consider the geometry of the 3D Heisenberg manifold and set $H$ to zero. This means that the ungauged action is simply
\bea\label{ugSh}
S=\int_{\S_2} \sfrac 12 g_{ij}(X)\dd X^i\w\star \dd X^j~, \eea
where the metric is 
\bea\label{mH}
\dd s^2=(\dd x^1)^2+(\dd x^2-x^1\dd x^3)^2 +(\dd x^3)^2,  \eea
in a particular coordinate system where the global 1-forms of the coframe and the corresponding dual vector 
fields are:
\bea\label{He}
&&e^a=\{\dd x^1, \dd x^2-x^1\dd x^3, \dd x^3\}~, \nn\\
&&\u_a=\{\partial_1, \partial_2, \partial_3+x^1\partial_2\}~,
\eea
and they satisfy
\be 
\dd e^2=-e^1\w e^3~ \quad\text{and} \quad  [\u_1,\u_3]=\u_2~.
\ee
Note that this is a case where the structure functions are constant.
We are going to use the vectors $\u_a$ to perform the T-duality and to this end 
we calculate the Lie derivative of the metric along them: 
\bea 
&&{\cal L}_{\u_1}g=-\dd x^2\otimes \dd x^3-\dd x^3\otimes \dd x^2+2x^1\dd x^3\otimes \dd x^3~,\nn\\
&&{\cal L}_{\u_2}g=0~,\nn\\
&&{\cal L}_{\u_3}g=\dd x^1\otimes \dd x^2+\dd x^2\otimes\dd x^1-x^1\dd x^1\otimes \dd x^3-x^1\dd x^3\otimes \dd x^1~.
\eea
We note that only $\u_2$ is a Killing vector. Recall that performing a standard T-duality transformation 
along this Killing direction one gets the well-known case of a 3-torus with $H$ flux. Moreover, 
it should be mentioned that the vector field $\partial_3$, which is not one of the $\u_a$ we considered, 
is also Killing and T-duality along this direction yields the case with non-geometric $Q^{23}_1$ flux. 
Here we take a different route.  

Following the general approach of section 2, we gauge the action (\ref{ugSh}) along all three vectors $\u_a$. 
In the present example we consider $\theta_a=0$ for simplicity.
Then the  gauged action reads as 
\bea\label{gSh}
S=\int_{\S_2} \sfrac 12 g_{ij}DX^i\w\star DX^j- \int_{\S_2} \dd\eta_a\wedge A^a+\int_{\S_2}\sfrac 12 C^c_{ab}\eta_cA^a\wedge A^b
 -\int_{\S_2}\omega^a_{bi}\eta_a A^b\wedge DX^i~,
 \eea
 where the metric is given above, $C^2_{13}=1$ is the only non-vanishing component of $C^a_{bc}$, and 
 \be 
 DX^1=\dd X^1-A^1~,\quad DX^2=\dd X^2-A^2-X^1A^3~,\quad DX^3=\dd X^3-A^3~.
 \ee
 The next step is to determine the $\omega^a_{bi}$ such that all the conditions and constraints are satisfied. 
 Performing this task, the constraints impose the only non-vanishing components of $\omega^a_{bi}$ to be 
 \be 
 \omega^2_{31}=-\omega^2_{13}=1~.
 \ee
 Then the system of equations is consistent and the action is gauge invariant.

It is interesting to note that in this example the connection 1-form $\omega$ is a very 
particular one. Using the basis $e^a$ of 1-forms it is a direct task to determine the connection $\Omega$ compatible with this basis by the tetrad postulate: 
\bea
D^\Omega e^a = de^a + \Omega^a_b \wedge e^b = 0 \;.
\eea
The only non-trivial relation $de^2 = -e^1\wedge e^3$ gives $\Omega^2_{31} = -\Omega^2_{13} = \tfrac{1}{2}$. Thus, the connection $\omega$ in the gauged sigma model is a constant multiple of the connection compatible with the orthonormal basis: $\omega = 2\Omega$.
Moreover, the curvature 2-form of $\omega^a_b$ vanishes, which is in accord with the gauge covariance of 
the field strength ${\cal F}^a$, as discussed in Section 2. It is reasonable to worry that then one can 
always set $\omega^a_b$ to zero by a suitable gauge transformation. However, once the vector fields 
$\u_a$ that implement the gauge symmetry are chosen, this is not possible any more. Of course, the geometry of the model might 
also possess sets of Killing vector fields, as in the present example, for which $\omega^a_b$ 
vanishes; it is however a legitimate 
choice \emph{not} to perform T-duality along them.

\subsection{Back to the original model}

In order to obtain the original model we should integrate out the Lagrange multiplier $\eta_a$. 
The corresponding equations of motion are
\bea 
\dd A^1&=&0~,\nn\\
\dd A^2+A^1\w A^3&=&A^3\w DX^1-A^1\w DX^3~,\nn\\
\dd A^3&=&0~.
\eea
Plugging them in the gauged action we obtain
\bea\label{gShoex}
S=\int_{\S_2} \sfrac 12 g_{ij}(X)DX^i\w\star DX^j~.\eea
At this stage we consider the gauge fixing procedure that was described in section 3. 
As mentioned there, one choice that fixes the gauge of $A^a$ is to set them to zero on-shell. 
Then the original ungauged action is recovered. 
More generally,
since $A^1$ and $A^3$ are closed we can choose a gauge where
\be 
A^1=\k_1\dd X^1 \quad\text{and}\quad A^3=\k_3\dd X^3~,
\ee 
for some real constants $\k_1, \k_3$. 
Then 
\bea 
DX^1&=&(1-\k_1)\dd X^1 := \dd Y^1~,
\nn\\
DX^3&=&(1-\k_3)\dd X^3 := \dd Y^3~,
\eea
where we defined the new coordinates $Y^1,Y^3$ as indicated by the last equations.
It remains to gauge fix $A^2$. In order to do this consistently first we note that the action \eqref{gShoex}
 is now written as 
\be 
\sfrac 12\int_{\S_2}\left( \dd Y^1\w\star \dd Y^1+ DX^2\w\star DX^2+ (1+\sfrac{(Y^1)^2}{(1-\k_1)^2})\dd Y^3\w\star \dd Y^3-\sfrac{2Y^1}{1-\k_1}DX^2\w\star \dd Y^3\right)~.
\ee
This allows us to determine a gauge choice for $A^2$ such that the original model is recovered:
\be \label{a2choice}
A^2=-(\k_1+\k_3-\k_1\k_3)X^1\dd X^3~.
\ee
Indeed then the action takes its final form in the new coordinate system (with $Y^2:= X^2$):
\be 
\int_{\S_2}\sfrac 12 g_{ij}(Y)\dd Y^i\w\star \dd Y^j~,
\ee
as desired. 
The remaining consistency check is that the chosen $A^2$ satisfies its equation of motion. 
Studying this equation we find that it is satisfied upon the choice \eqref{a2choice} without further restrictions.

\subsection{The dual model}

Finally let us obtain the dual model, which is the most interesting instance of our analysis. 
First we have to integrate out the gauge fields. Thus we vary the action (\ref{gSh}) 
with respect to $A^a$ and obtain the equations of motion 
\bea 
A^1&=&\dd X^1-\star(\dd \eta_1-\eta_2 A^3+\eta_2\dd X^3)~,\nn\\
A^2&=&\dd X^2-X^1\dd X^3-\star\dd\eta_2~,\nn\\
A^3&=&\dd X^3-\star(\dd\eta_3+\eta_2 A^1-\eta_2\dd X^1)~.
\eea
These equations are coupled but we can easily decouple them and find
\bea\label{Asolh}
A^1&=&\dd X^1-\sfrac{\eta_2}{1+(\eta_2)^2}\dd\eta_3-\sfrac 1{1+(\eta_2)^2}\ast\dd\eta_1~,\nn\\
A^2&=&\dd X^2-X^1\dd X^3-\ast\dd \eta_2~,\nn\\
A^3&=&\dd X^3+\sfrac{\eta_2}{1+(\eta_2)^2}\dd\eta_1-\sfrac 1{1+(\eta_2)^2}\ast\dd\eta_3~,
\label{eomsh} \eea
Inserting the equations of motion in the action, a tedious calculation leads to the simple result
\bea\label{dualh}
S_{\text{dual}}&=&\frac 12\int_{\S} \left(\dd \eta_2\w\ast \dd \eta_2+
\frac{1}{1+(\eta_2)^2}(\dd\eta_1\w\ast \dd\eta_1+
\dd \eta_3\w\ast \dd\eta_3)\right. \nn\\ &&\qquad~~ \left.+\frac{2\eta_2}{1+(\eta_2)^2}\dd\eta_1\w \dd\eta_3\right)~,
\eea
Let us comment on this result. The action of the dual model looks precisely like the so-called $Q$-flux background or T-fold. Indeed 
the background fields in the basis $\{\dd\eta_a\}$ are 
\bea
g&=&\dd\eta_2\otimes \dd\eta_2+\frac{1}{1+(\eta_2)^2}(\dd\eta_1\otimes \dd\eta_1+
\dd\eta_3\otimes \dd\eta_3)~,\\
B&=&\frac{\eta_2}{1+(\eta_2)^2}\dd\eta_1\w \dd\eta_3~,
\eea 
which are globally ill-defined if the direction $\eta_2$ is assumed compact, unless O(3,3) transformations 
are allowed as transition functions in the patching of the background fields \cite{tfold1,tfold2}. 
This then corresponds to a short cut for the following chain of standard, isometric T-dualities
\bea 
C^2_{13} \quad \overset{T_2}\to \quad H_{123} \quad \overset{T_1}\to \quad C^1_{23} \quad 
\overset{T_3}\to \quad Q^{13}_2~. \label{chain}
\eea 
This can be depicted in terms of the following commutative diagram of isometric/non-isometric T-dualities:
\bea 
 \xymatrix{  H_{123}\ar@(ul,ur)^{\d B} \ar[r]^{\text{T}^{\text{iso}}_{\partial_1}}  &C^1_{23} \ar[d]^{\text{T}^{\text{iso}}_{\partial_3}}   \\
  \:\, C^2_{13} \ar[u]^{\text{T}^{\text{iso}}_{\partial_2}}  \ar[r]^{\text{T}^{\text{non-iso}}_{\u_a}} & Q^{13}_{2}}
\eea
The commutativity of this diagram is interesting, since the Killing vector fields for the isometric 
route are not all the same with the ones in the non-isometric route---in particular the third one is 
$\partial_3$ and $\u_3=\partial_3+x^1\partial_2$ respectively. Thus in this example we encounter a 
 non-isometric short cut for an isometric duality chain, which is now possible in one step and in a non-Abelian fashion. Of course the real challenge is
to perform a T-duality for a case that cannot be handled with standard methods at all. We leave this for 
a future publication. Finally it is worth mentioning the obvious fact that the original and the dual coordinates for this example are fully disentangled.

\section{A class of examples}

The toy example we studied in the previous section provides an existence proof for non-trivial cases 
where the present formulation applies. 
Furthermore it indicates that there exists a considerably large class of additional examples based 
on nilmanifolds. Here we formulate non-isometric T-duality for an arbitrary step 2 nilmanifold in any 
dimension. 

In all cases we are working with pure geometries, namely we set $H=0$.
We write the ungauged action in the form 
\be 
S=\int_{\S_2}\sfrac 12 \d_{ab}e^a\w\star e^b~,
\ee
where $e^a$ is a global coframe. In a coordinate basis where 
$e^a=e^a_{i}\dd x^i$, where $e^a_i$ are the (inverse) vielbeins, the metric takes the form
\be 
g=\d_{ab}e^a_{i}e^b_j\dd x^i\otimes\dd x^j~,
\ee
and the action becomes the same as in the previous sections. 
The set of vector fields that we use for T-duality is the one given by the dual frame, i.e.
\be 
\langle \u_a,e^b\rangle=\d_a^b~.
\ee
A useful relation is
\be 
{\cal L}_{\u_a}e^b=-C_{ac}^be^c~,
\ee 
where $C_{ab}^c$ are the structure constants of the algebra of vector fields, 
\be 
[\u_a,\u_b]=C^a_{bc}\u_c~,
\ee
which also appear in the Maurer-Cartan equations
\be 
\dd e^a=-\sfrac 12 C^a_{bc}e^b\w e^c~.
\ee
Then it is simple to compute the Lie derivative of the metric along these vector fields. 
This yields the result
\be 
{\cal L}_{\u_a}g=-\sum_cC^c_{ab}e^b\vee e^c~.
\ee
On the other hand, we have to solve the condition
\be 
{\cal L}_{\u_a}g=\o_a^b\vee\iota_{\u_b}g~.
\ee 
This means that
\be 
\sum_c \o^c_{ab}e^b\vee e^c=-\sum_cC^c_{ab}e^b\vee e^c~.
\ee 
This equation is solved by 
\be 
\o^c_{ab}=-C^c_{ab}~,
\ee 
which is consistent with the results of our previous example.
Moreover, assuming again $\theta_a=0$, all the constraints imposed by 
gauge invariance are satisfied. Then the gauged action is 
\bea  
S&=&
\int_2 \left(\sfrac 12 \d_{ab}E^a\w\star E^b- \dd\eta_a\wedge A^a+\sfrac 12 C^c_{ab}\eta_cA^a\wedge A^b -
\omega^c_{ab}\eta_c A^a\wedge E^b\right)
\nn\\
&=&\int_2\left( \sfrac 12 \d_{ab}E^a\w\star E^b- \dd\eta_a\wedge A^a+\sfrac 12 C^c_{ab}\eta_cA^a\wedge (A^b+2E^b)\right)~,
\eea 
where $E^a=e^a-A^a$.

As before, the original model is recovered by integrating out the Lagrange multipliers $\eta_a$. 
This leads to the equations of motion
\be 
\dd A^a+\sfrac 12 C^a_{bc}A^b\w(A^c+2E^c)=0~,
\ee
which are then inserted to the gauged action and yield
\be \label{nilmodel}
S=\int_{\S_2}\sfrac 12 \d_{ab}E^a\w\star E^b~.
\ee
Now we have to follow a gauge fixing procedure. 
We make the general Ansatz 
\be 
A^a=\k^a_{b}e^b+\l^c_d C^a_{bc}X^b e^d~,
\ee
for sets of real constants $\k$ and $\l$. 
In order to proceed, we have to use some properties of step 2 nilmanifolds. 
To this end we consider the splitting of the indices $a=(a_0,\bar a)$ such that 
$C^{a_0}_{bc}=0$ and $C^{\bar{a}}_{bc}\ne 0$. This is always possible because by definition there is always a subset 
of vanishing structure constants for
nilmanifolds, due to nilpotency. Using the fundamental step 2 relation 
\beq
C^a_{bc}C^{c}_{de}=0~,
\eeq
which is true even without summation in the index $c$,
it is evident that $C^a_{\bar{b}c}=0$ for all indices with a bar.
Under this splitting, the Ansatz for the gauge field becomes
\bea 
A^{a_0}&=&\k^{a_0}_be^b~,\nn\\
A^{\bar{a}}&=&\k^{\bar{a}}_be^b+\l^{c_0}_dC^{\bar{a}}_{b_0c_0}X^{b_0}e^d~.
\eea
First let us choose $\k^{\bar{a}}_b=0, \k^{a_0}_{\bar{b}}=0$ and $\l^{a_0}_{\bar{b}}=0$. 
Then we compute
\be 
\dd E^{a_0}=0~
\ee
identically, and 
\be 
\dd E^{\bar{a}}=-\sfrac 12 C^{\bar{a}}_{b_0c_0}E^{b_0}\w E^{c_0}~,
\ee 
provided that 
\be 
C^{\bar{a}}_{c_0[b_0}\l^{c_0}_{d_0]}=C^{\bar{a}}_{p_0c_0}\k^{c_0}_{[d_0}(1-\sfrac 12 
\k)^{p_0}_{b_0]}~.\label{conditionorig}
\ee 
Then the Ansatz for the gauge fields
leads to
\be 
\dd E^a=-\sfrac 12 C^{a}_{bc}E^b\w E^c~.
\ee
This means that the action \eqref{nilmodel} is precisely the action of the original model. 
As before, it remains to guarantee that the gauge field satisfies the equations of motion. 
It is straightforward to check that \eqref{conditionorig} is enough to attain this and no further 
restrictions are imposed.
Let us note that for step 2 nilmanifolds of dimension  $d\le 7$, which are fully classified and anyway they 
constitute the interesting cases for our purposes, the number of unknowns is larger than the number of equations that constrain them. Therefore the procedure works.
Above 7 dimensions we do not make a general claim since there is no classification of nilmanifolds \cite{goze}.

Now let us integrate out the gauge fields. The corresponding equations of motion are 
\be 
(\d_{ab}-C^c_{ab}\eta_c \star)A^b=(\d_{ab}-C^c_{ab}\eta_c\star)e^b-\star\dd\eta_a~.
\ee
In order to insert this equation in the action we have to determine $A^a$. 
Following the same procedure as in the general case, the result is
\be 
A^a=e^a-\d^{ad}C^c_{dp}\eta_c S^{pb}\dd\eta_b- S^{ab}\star\dd\eta_b~,
\ee
where we defined 
\be 
S^{ab}=(\d_{ab}+C^d_{bp}C^c_{aq}\d^{pq}\eta_d\eta_c)^{-1}~.
\ee
Inserting this expression in the action we obtain
\bea 
S=\int_{\S_2}\biggl(\sfrac 12 S^{ab}\dd\eta_a\w\star\dd\eta_b-\sfrac 12 \d^{bm}C^d_{mp}\eta_d S^{pa}\dd\eta_a\w \dd\eta_b\biggl)~.
\eea
This is a generalization for any 
step 2 nilmanifold of the dual model we found in the example of the previous section. In particular we observe also in this more general situation that the original and the dual coordinates disentangle.

\section{Conclusions}

The range of validity of the Buscher rules for the T-duality of string background fields is 
limited to the case where isometries are present and additional invariance conditions are imposed.
In this paper we used a recent idea about gauge symmetries of 2D sigma models without corresponding 
global symmetry \cite{Kotov:2014iha} to study T-duality in a more general setting\footnote{Additional symmetries appearing in 2D sigma models and their role in the context of T-duality are under investigation \cite{SeveraStrobl}.}. 
In particular we were able to identify the conditions and constraints that guarantee that a 
bosonic sigma model with a metric and $B$-field is gauge invariant under an extended set of gauge transformations 
even when one does not have isometries at hand. 
All these conditions are milder than their counterparts in the isometric case. The next step was 
to follow the standard procedure of Buscher in this non-standard setting. Integrating out 
the Lagrange multipliers from the gauged action and gauge fixing lead back to the original ungauged 
model. On the other hand, integrating out the gauge fields from the action yields a dual model which was 
precisely identified.

Since several constraints are involved in the formulation, it is natural to worry whether there is any 
room for non-trivial applications. As a proof of existence, we studied a particular geometry which 
is often considered in string theory as a useful toy model. This geometry corresponds to the 3D 
Heisenberg nilmanifold and carries no $H$ flux. In that case we determined a solution of all 
conditions and constraints that allows to T-dualize along vector fields that are not Killing. 
This led to a dual model being identical to the T-fold geometry that is associated to non-geometric Q-flux, 
as discussed for example in Refs. 
\cite{tfold1,tfold2} from a different perspective. Additionally we showed that this is not an isolated case, but in fact all 
step 2 nilmanifolds in dimensions up to and including 7 provide a class of working examples.

Although the above results are encouraging, there are certain limitations in their scope as presented in this paper, and it is useful to mention some of them. First of all, the dilaton was ignored. 
The transformation of the dilaton involves an 1-loop computation since the corresponding coupling appears at first order in $\alpha'$, which should be examined. Moreover, we did not discuss at all the 
potential equivalence of the 
two dual models as conformal field theories, which is true in the standard case 
and requires a careful consideration for global issues of the procedure \cite{RocekVerlinde}. 
Furthermore a better understanding of the underlying geometric structures is due. 
In the cases we examined we found a mixing of the original coordinates and the would-be dual coordinates, 
which is indicative of doubled formulations, such as the doubled sigma models considered by 
Hull \cite{Hull1} or the ones recently studied in \cite{CJL}.
Last but not least, it would be very interesting to apply this formalism in the case of the triple T-dual 
of a torus with $H$ flux, or equivalently to the T-dual of the $Q$ flux background where no 
isometry is available (recent attempts to understand this problem include \cite{Blumenhagen:2011ph, Andriot:2012vb, Bakas:2015gia}), and even more so in cases of true string backgrounds.  
We will report on these and other issues in future publications.

 \newpage
 \paragraph{Acknowledgements.} We are deeply grateful to Thomas Strobl for explaining to us 
 numerous ideas related to the present work and for sharing with us his insights on the topic. 
 We would also like to thank Chris Hull, George Papadopoulos, Soo-Jong Rey, Pavol $\mathrm{\check S}$evera and Jim Stasheff for questions 
 and related discussions. 
 The main part of this work was done during the CERN-CKC TH Institute on Duality Symmetries in String and M-Theories; A.C. and L.J. would like to thank the organisers and acknowledge financial support from CERN. The work of L.J. was supported in part  by the  Alexander von Humboldt Foundation and by Croatian Science Foundation under the project IP-2014-09-3258.


\begin{thebibliography}{99}
 \addtolength{\itemsep}{-4pt}  
  
\bibitem{Polchinski:2014mva}
  J.~Polchinski,
  ``Dualities of Fields and Strings,''
  arXiv:1412.5704 [hep-th].
  
  
\bibitem{Giveon:1994fu}
  A.~Giveon, M.~Porrati and E.~Rabinovici,
  Phys.\ Rept.\  {\bf 244} (1994) 77
  [hep-th/9401139].
  
  \bibitem{Buscher1}
  T.~H.~Buscher,
  Phys.\ Lett.\ B {\bf 194} (1987) 59.
  
\bibitem{Buscher2}
  T.~H.~Buscher,
  Phys.\ Lett.\ B {\bf 201} (1988) 466.
  
\bibitem{RocekVerlinde}
  M.~Ro$\mathrm{\check c}$ek and E.~P.~Verlinde,
  Nucl.\ Phys.\ B {\bf 373} (1992) 630
  [hep-th/9110053].
  
\bibitem{HullSpence1}
  C.~M.~Hull and B.~J.~Spence,
  Phys.\ Lett.\ B {\bf 232} (1989) 204.
  
\bibitem{HullSpence2}
  C.~M.~Hull and B.~J.~Spence,
  Nucl.\ Phys.\ B {\bf 353} (1991) 379.
  
\bibitem{Hull}
  C.~M.~Hull,
  JHEP {\bf 0710} (2007) 057
  [hep-th/0604178].
  
\bibitem{delaOssa}
  X.~C.~de la Ossa and F.~Quevedo,
  Nucl.\ Phys.\ B {\bf 403} (1993) 377
  [hep-th/9210021].
  
\bibitem{Alvarez1}
  E.~Alvarez, L.~Alvarez-Gaum\'e, J.~L.~F.~Barb\'on and Y.~Lozano,
  Nucl.\ Phys.\ B {\bf 415} (1994) 71
  [hep-th/9309039].
  
\bibitem{Alvarez2}
  E.~Alvarez, L.~Alvarez-Gaum\'e and Y.~Lozano,
  Nucl.\ Phys.\ B {\bf 424} (1994) 155 \newline
  [hep-th/9403155].
  
\bibitem{Plauschinn1}
  E.~Plauschinn,
  JHEP {\bf 1401} (2014) 131
  [arXiv:1310.4194 [hep-th]].
  
   \bibitem{Plauschinn2}
  E.~Plauschinn,
  Nucl.\ Phys.\ B {\bf 893} (2015) 257
  [arXiv:1408.1715 [hep-th]].
  
\bibitem{Kotov:2014iha}
  A.~Kotov and T.~Strobl,
  J.\ Geom.\ Phys.\  {\bf 99} (2016) 184
  [arXiv:1403.8119 [hep-th]].
  
\bibitem{Strobl:2004im}
  T.~Strobl,
  Phys.\ Rev.\ Lett.\  {\bf 93} (2004) 211601
  [hep-th/0406215].


\bibitem{Bojowald:2004wu}
  M.~Bojowald, A.~Kotov and T.~Strobl,
  J.\ Geom.\ Phys.\  {\bf 54} (2005) 400
  \newline [math/0406445 [math-dg]].
  
\bibitem{Strobl1}
  A.~Kotov and T.~Strobl,
  ``Generalizing Geometry - Algebroids and Sigma Models,''
  arXiv:1004.0632 [hep-th].
  
\bibitem{Strobl2}
  V.~Salnikov and T.~Strobl,
  JHEP {\bf 1311} (2013) 110
  [arXiv:1311.7116 [math-ph]].
  
\bibitem{Strobl3}
  A.~Kotov, V.~Salnikov and T.~Strobl,
  JHEP {\bf 1408} (2014) 021
 \newline [arXiv:1407.5439 [hep-th]].

  \bibitem{Mayer}
  C.~Mayer and T.~Strobl,
  J.\ Geom.\ Phys.\  {\bf 59} (2009) 1613
  [arXiv:0908.3161 [hep-th]].



\bibitem{Klimcik:1995ux}
  C.~Klim$\mathrm{\check c}$\'ik  and P.~$\mathrm{\check S}$evera,
  Phys.\ Lett.\ B {\bf 351} (1995) 455
  [hep-th/9502122].

\bibitem{Klimcik:1995jn}
  C.~Klim$\mathrm{\check c}$\'ik,
  Nucl.\ Phys.\ Proc.\ Suppl.\  {\bf 46} (1996) 116
  [hep-th/9509095].

\bibitem{Klimcik:1995dy}
  C.~Klim$\mathrm{\check c}$\'ik and P.~$\mathrm{\check S}$evera,
  Phys.\ Lett.\ B {\bf 372} (1996) 65
  [hep-th/9512040].

\bibitem{Sfetsos:1997pi}
  K.~Sfetsos,
  Nucl.\ Phys.\ B {\bf 517} (1998) 549
  [hep-th/9710163].

\bibitem{Davidovic:2015mpa}
  L.~Davidovi\'c and B.~Sazdovi\'c,
  JHEP {\bf 1511} (2015) 119
  [arXiv:1505.07301 [hep-th]].


 \bibitem{Strobltoappear}
  A.~Kotov and T.~Strobl,
  Phys.\ Rev.\ D {\bf 92} (2015) 085032.


 \bibitem{CDJS}
  A.~Chatzistavrakidis, A.~Deser, L.~Jonke and T.~Strobl, ``Gauging of Foliations and Universal 
  Gauge Theory for Bosonic Strings,''  in preparation.

\bibitem{tfold1}
  C.~M.~Hull and R.~A.~Reid-Edwards,
  JHEP {\bf 0909} (2009) 014
  [arXiv:0902.4032 [hep-th]].
  
\bibitem{tfold2}
  R.~A.~Reid-Edwards,
  JHEP {\bf 0906} (2009) 085
  [arXiv:0904.0380 [hep-th]].

 


 
 \bibitem{goze}
M.~Goze and Y.~Khakimdjanov,
\textit{Nilpotent Lie Algebras,} Mathematics and its Applications, \textbf{361}, Kluwer Academic Publishers Group, Dordrecht, 1996. 
  
\bibitem{SeveraStrobl}
P.~$\mathrm{\check S}$evera and T.~Strobl, ``T-duality without symmetries and DFT without coordinates'', work in progress.

\bibitem{Hull1}
  C.~M.~Hull,
  JHEP {\bf 0510} (2005) 065
  [hep-th/0406102].

\bibitem{CJL}
A.~Chatzistavrakidis, L.~Jonke and O.~Lechtenfeld,
  JHEP {\bf 1511} (2015) 182
  [arXiv:1505.05457 [hep-th]].

\bibitem{Blumenhagen:2011ph} 
  R.~Blumenhagen, A.~Deser, D.~L\"ust, E.~Plauschinn and F.~Rennecke,
  J.\ Phys.\ A {\bf 44}, 385401 (2011)
  [arXiv:1106.0316 [hep-th]].

\bibitem{Andriot:2012vb}
  D.~Andriot, M.~Larfors, D.~L\"ust and P.~Patalong,
  JHEP {\bf 1306} (2013) 021
  [arXiv:1211.6437 [hep-th]].

\bibitem{Bakas:2015gia}
  I.~Bakas and D.~L\"ust,
  ``T-duality, Quotients and Currents for Non-Geometric Closed Strings,''
  arXiv:1505.04004 [hep-th].

 

  
 \end{thebibliography}
\end{document}